\documentclass[10pt,a4paper]{article}
\usepackage[left=2.5cm,top=3cm,right=2.5cm,bottom=3cm]{geometry}

\usepackage{amsmath} 

\usepackage{graphicx}
\usepackage[small,bf]{caption} 
\usepackage{subfig}
\usepackage{color}
\usepackage[colorlinks=true,allcolors=black,urlcolor=blue,]{hyperref}

\usepackage[round]{natbib}

\begin{document}

\begin{center}
\Large\textbf{Biologically inspired force enhancement \\ for maritime propulsion and maneuvering}\\[1em]
\large\textbf{Gabriel D. Weymouth},
University of Southampton, UK, \\
\href{mailto:G.D.Weymouth@soton.ac.uk}{G.D.Weymouth@soton.ac.uk}\\[3em]
\end{center}

\thispagestyle{empty}
\pagestyle{empty}

\begin{abstract}
The move to high performance applications greatly increases the demand to produce large instantaneous fluid forces for high-speed maneuvering and improved power efficiency for sustained propulsion. Animals achieve remarkable feats of maneuvering and efficiency by changing their body shape to generate unsteady fluid forces. Inspired by this, we have studied a range of immersed bodies which drastically change their shape to produce fluid forces. These include relatively simple shape-changes, such as quickly changing the angle of attack of a foil to induce emergency stops and the use of tandem flapping foils to generate three times the average propulsive force of a single flapping foil. They also include more unconventional shape-changes such as high-speed retracting foil sections to power roll and dive maneuvers and the use of soft robotics to rapidly shrink the frontal area of an ellipsoid to power 68\% efficient fast-start maneuvers or even completely cancel the drag force with 91\% quasi-propulsive efficiency. These systems have been investigated with analytics, experimental measurements and immersed-boundary numerical simulations.

\end{abstract}

\section{Introduction}

With the expansion of human activities in the oceans towards more extreme environments, state-of-the-art maritime technologies have progressively become less suited at coping with the increased degree of complexity of their missions. As an example, the offshore oil industry is more and more involved in operating in deeper waters and need to acquire baseline and on-going surveys throughout the life history of submerged infrastructures and their interaction with the surrounding ecosystems. Currently, operations of this kind rely heavily on expensive and slow human divers because traditional robots are not as well suited to acquiring in-situ measurements in very close proximity to submerged structures or living organisms.

Aerial and marine animal achieve remarkable feats of maneuvering and efficiency by changing their body shape to generate unsteady fluid forces. For example, birds execute precise maneuvers, such as banking, braking, takeoff and landing, all with minimal power expended \citep{Provini2014}. This is in stark contrast to current ``flight-type'' marine and aerial vehicles with fixed wings which have a fixed minimum operating speed and slow response time, or ``hover-type'' vehicles with multiple thrusters which have limited mission lives due to their inefficiency. 

Starting with the seminal work of \cite{Lighthill1960}, which mathematically formulated how fish produce large forces and high efficiency with undulatory motion, there has been significant research in studying shape-changing unsteady biological flows and exploiting them in maritime engineering designs. While fish swimming itself has now been well studied \citep{Triantafyllou2000} and applied to small robotic vehicles \citep{Triantafyllou1994}, the mechanical complexities make it difficult to adapt fish-propulsion to broader applications. In this manuscript, we review some recent work on biologically inspired mechanisms which generate strong forces, are highly efficient, and are achieved with relatively simple actuation methods, all of which makes them potentially well-suited to maritime applications.

\section{Heaving and pitching foils}\label{sec:flap}

The first biologically inspired force-producing device was certainly a flapping wing, dating at least as far back as Da Vinci ca. 1485 \citep{Mccurdy1941}. Modern research has revitalized this concept, showing that lifting surfaces which are actuated to dynamically heave and/or pitch have potential advantages over either fixed lifting surfaces or standard propellers. Studies on the thrust forces generated by an oscillating foil have shown the potential for impressive thrust coefficients (maximum of $C_T=2.4$) and efficiency regions of 50-60\% \citep{Read2003}. It has also been shown that an oscillating foil can be used to manipulate incoming vorticity for energy extraction, with efficiencies at and above 45\% \citep{Simpson2008}. However, there are a wide range in observed efficiencies and force magnitudes, and these parameters vary with oscillation type, planform and flexibility of the foil. This section reviews two studies on actuated rigid foils which demonstrate large force production at high efficiency levels with simple kinematics.

\subsection{Tandem flapping foils to balance forces and utilize wake energy}

A fundamental issue with implementing a flapping foil as a marine propulsor on an otherwise conventional ship or underwater vehicle is the large variation in thrust and side force. Additionally, propulsive efficiency in the range of 50-60\% is not optimum, indicating that mechanical power is being wasted in energizing the wake. A recent study by \cite{Epps2016} investigated the use of tandem flapping foils to mitigate the unbalanced forces and potentially increase efficiency by utilizing energy in the wake of the forward foil. 

\begin{figure}
	\centering
	\subfloat[Single]{
		\includegraphics[width=0.5\textwidth]
		{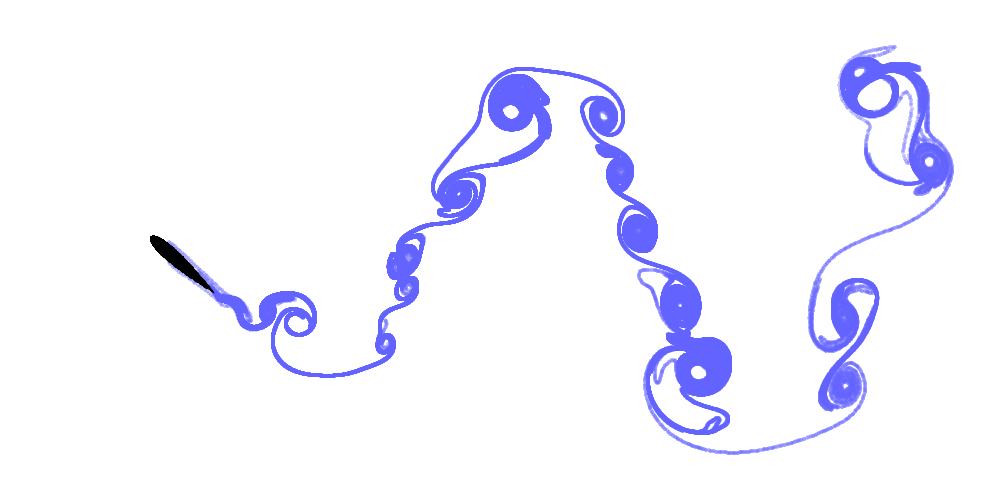}
		\label{fig:single}}
	\subfloat[Tandem]{
		\includegraphics[width=0.5\textwidth]
		{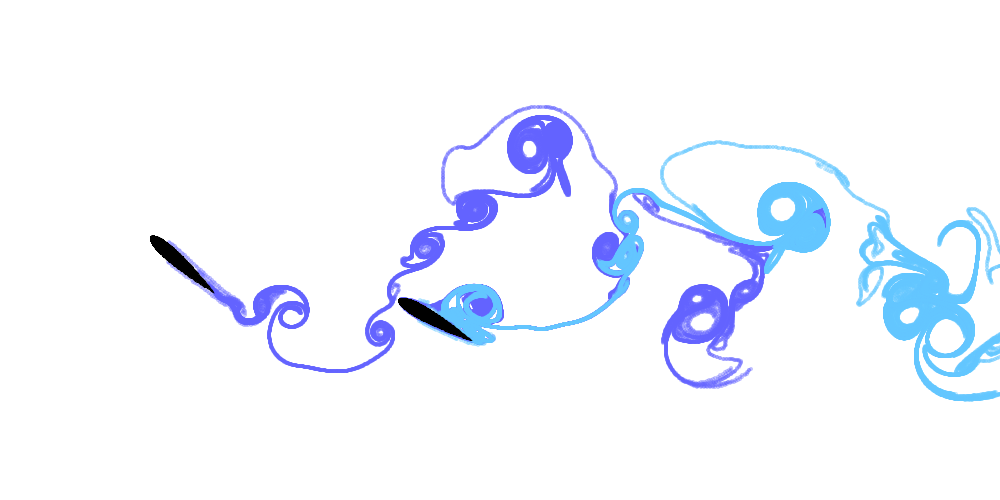}
		\label{fig:tandem}}
	\caption{Simulated streaklines for the two-dimensional flow past a single flapping foil and tandem flapping foils. The streaklines are visualized by continuously releasing tracer particles on either side of the foil at the quarter-chord. The tandem case has phase lag $\phi=1.75\pi$, and spacing $s=2c$.}
	\label{fig:foils}
\end{figure}

In this study, the foils undergo prescribed harmonic heave $h$ and pitch $\theta$, defined as
\begin{align}
& h_f(t) = c \sin(\omega t), \quad h_b(t) = c\sin(\omega t+\phi) \\
& \theta_f(t) = \frac \pi 4 \cos(\omega t), \quad \theta_b(t) = \frac \pi 4 \cos(\omega t+\phi)
\end{align}
where $c$ is the chord length, $\omega$ is the flapping frequency and $\phi$ is the phase lag between the foils, and the $f,b$ subscripts refer to the front and back foils respectively. The frequency is set to achieve a Strouhal number of $St = 4\pi\omega c / U = 0.4$, known to be at the upper end of the range resulting in high thrust for a single foil \citep{Read2003}. The flow speed $U$ is set to achieve a Reynolds number of $Re=Uc/\nu=10^4$.

This flow was studied using the Lily Pad computational fluid dynamics software. As discussed in \cite{Weymouth2015b}, Lily Pad is a two-dimensional Cartesian-grid flow solver that uses the Boundary Data Immersion Method  \citep[see][]{Maertens2015} and has been extensively validated for unsteady fluid-body interaction problems. For these simulations, a grid spacing of $h=c/64$ and a domain size of $16c$ x $8c$ was used.

Figure~\ref{fig:foils} shows a set of Lily Pad results for the flow around single and tandem flapping foils. Streaklines in Figure~\ref{fig:single} show that the characteristic reverse K\'arm\'an street has formed, accelerating the flow behind the single foil. Figure~\ref{fig:tandem} shows a set of streaklines for a tandem case where the leading edge of the back foil is spaced $s=2c$ behind the trailing edge of the front foil and the motion is lagged by $\phi=1.75\pi$. The wake in the tandem case has narrowed and lengthened compared to the single foil case, indicating greater speed and possibly efficiency.

\begin{figure}
	\centering
	\subfloat[Thrust]{
		\includegraphics[width=0.3\textwidth]{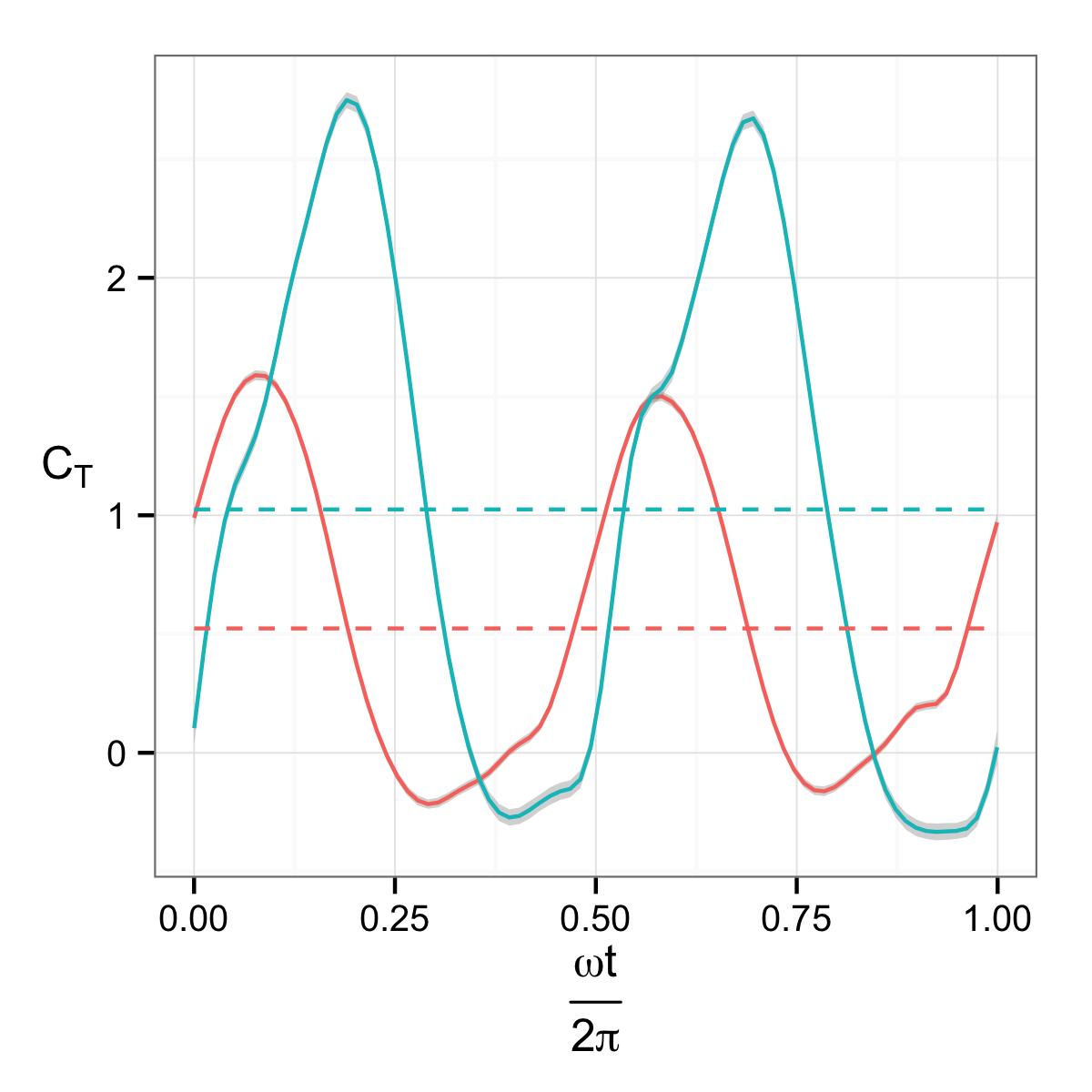}
		}
	\subfloat[Lift]{
		\includegraphics[width=0.3\textwidth]{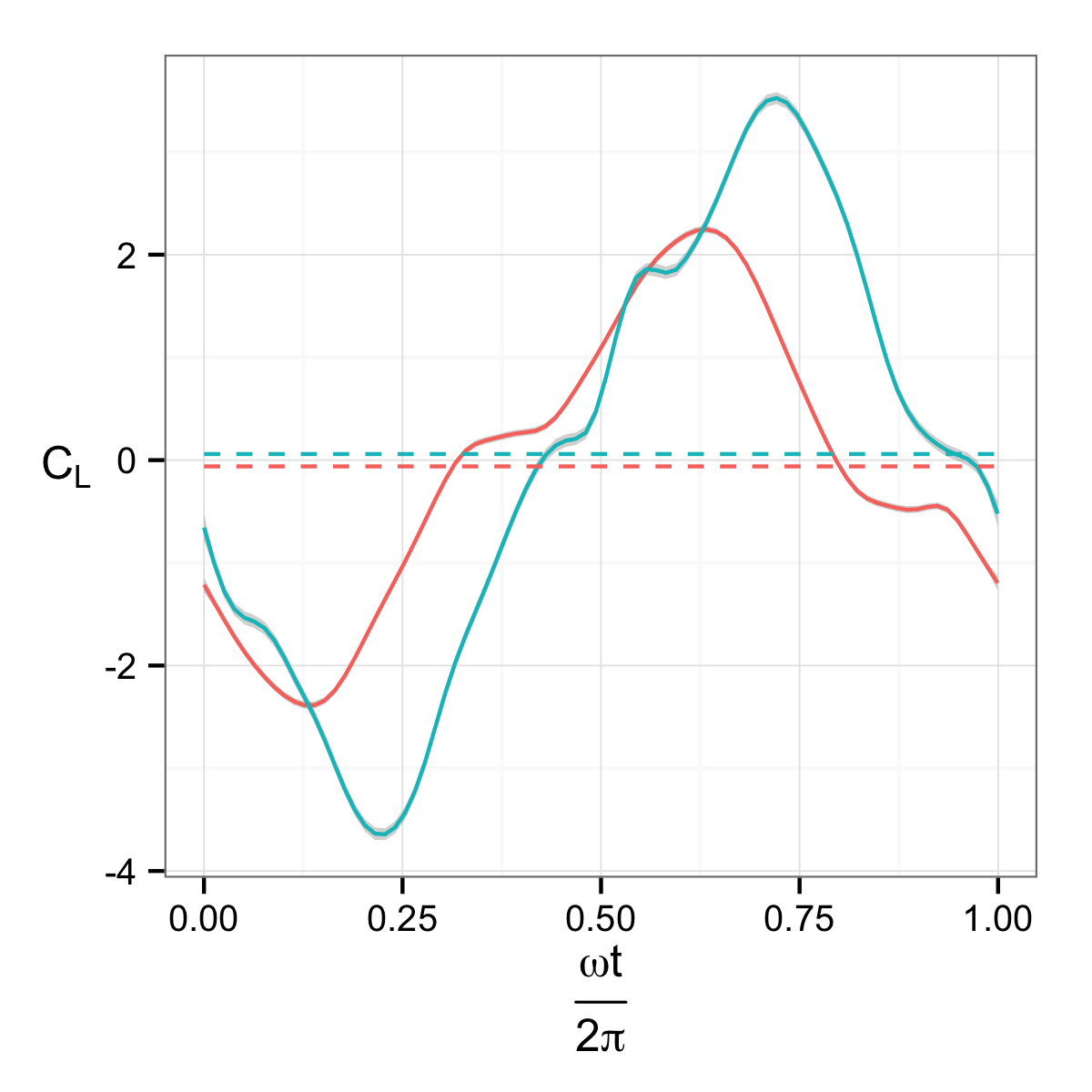}
		}
	\subfloat[Power]{
		\includegraphics[width=0.3\textwidth]{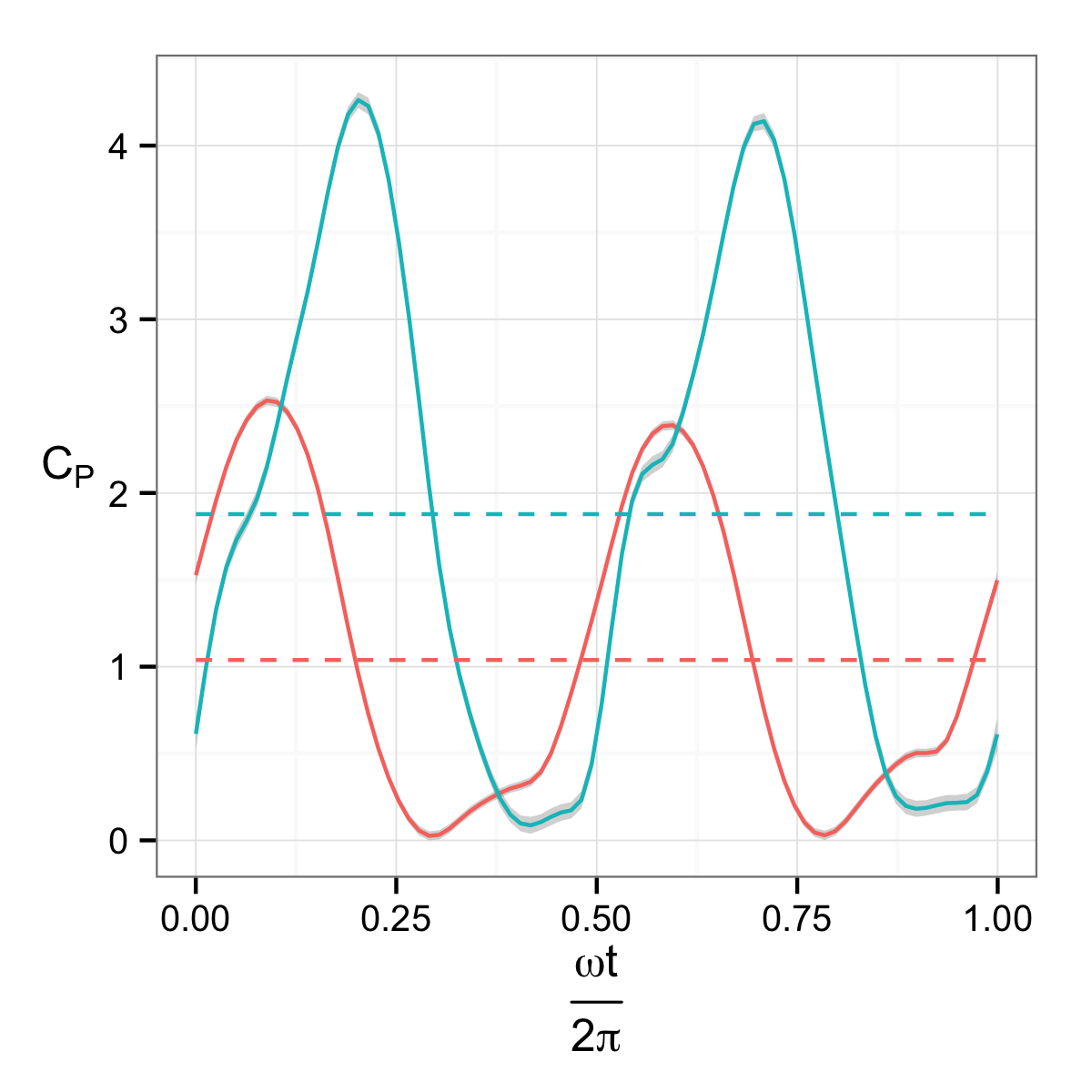}
		}
	\caption{Performance coefficients for the tandem foils shown in Figure~\ref{fig:tandem}; \textcolor[rgb]{0.8,0,0}{front foil}, \textcolor[rgb]{0,0.5,0.5}{back foil}; dashed lines are the mean values over the cycle.}
	\label{fig:foilsforces}
\end{figure}

A set of performance metrics are shown in Figure~\ref{fig:foilsforces}. The thrust $T$ and lift $L$ are defined as the integrated fluid force inline with and perpendicular to the oncoming flow, as usual. The general equation for the power transferred from the body to the fluid is
\begin{equation}\label{eq:p}
P = \oint_S \left(\vec f(s,t) \cdot \vec u(s,t)\right)\ ds
\end{equation}
where $\vec f$ is the local fluid force per unit area on the body surface, $\vec u$ is the local body surface velocity, $\oint_S\ ds$ is an integral over the body surface. This formula automatically accounts for both the pitch and heave motion and is also valid for the flexible and deforming bodies used in the next sections.

Another key performance metric is the efficiency, which is \textit{the rate of useful work done per unit power consumed}. As such, the hydrodynamic efficiency of a propulsive actuator operating at a steady forward speed is simply 
\begin{equation}\label{eq:etaT}
\eta_{t} = \frac{TU}{P}
\end{equation}
where $TU$ is the rate of work done in the inline direction.

The results in Figure~\ref{fig:foilsforces} are for the tandem case, but the performance of the front foil is essentially independent of the back foil for $s>c$. The front foil results compare well to those presented in the literature for single flapping foils, with a mean thrust coefficient of $C_{T,f}= T_f/(\frac 12 \rho U^2 c) = 0.52$, mean lift of zero, and mean power coefficient of $ C_{P,f}=P_f/(\frac 12 \rho U^3 c) = 1.04$. Therefore the efficiency for this simple choice of kinematics is 50\%.

The back foil undergoes the same motion as the front, but operates in its wake, which significantly changes the response. Most noticeable is the large increase this enables in the back foil thrust, $ C_{T,b}=1.02$, twice the value of the front foil. In other words, adding a second foil has not doubled the total thrust, but instead tripled it. This is due to the positive wake interference of the two foils. Negative interference is also possible, and \cite{Epps2016} develops a relationship between the spacing and phase to characterize this interference. 

In addition, the peak forces on the hind foil are phase shifted by $\phi$ relative to the front foil. By properly setting the spacing and phase, \cite{Epps2016} shows that a tandem foil propulsion system would be capable of greatly reducing the variation in the thrust force compared to a single flapping foil. It is also possible to reduce the variation in lift, but because the thrust peaks are twice as frequent, two foils cannot perfectly cancel both thrust and lift variation.

Finally, the increased thrust on the back foil shown in Figure~\ref{fig:foilsforces} does require increased power, but not disproportionally. In fact, the efficiency of the tandem foil system overall is $\eta_t=53\%$, slightly better than that of the front foil alone.

\subsection{Rapid pitch-up for impulsive stopping force}

One of the most striking advantages of flying animals over fixed-wing aircraft is their ability to come to a complete and controlled landing in only a few body lengths; even large gliding birds such as an eagle \citep{Carruthers2007}. Like aircraft, flight-type underwater vehicles have a minimum operating speed to maintain their depth, and because maritime vessels are proportionally much heavier than aircraft, they are even slower to stop. \cite{Polet2015} studied a simple model of wing kinematics during perching and found that very large dynamic lift and drag forces are produced - and these forces could potentially be utilized to impulsively stop heavy and streamlined maritime vehicles. 

\cite{Polet2015} focused on one key kinematic characteristic of bird perching, the rapid increase in pitch of the wings during deceleration. Lily Pad simulations ($Re=2000$) and experiments ($Re=22000$) were performed in which the foil speed and pitch angled were varied during the maneuver as 
\begin{align}
U(t) &= U_0 (1-t^*) \label{eq:stop}\\
\theta(t) &= \theta_{final} \left(t^*-\frac{\sin(2\pi t^*)}{2\pi}\right) \label{eq:turn}
\end{align}
 where $U_0$ is the initial velocity, $\theta_{final}=\frac \pi 2$ is the final pitch position, and $t^*= t/\tau$ is time scaled by the period of the maneuver $\tau$ up to $\theta=\pi/2$. A NACA0012 foil section was used and the center of rotation was set $c/6$ from the leading edge. 
 
We quantify the impulsiveness of the maneuver using the shape-change number
\begin{equation}\label{eq:Xi}
\Xi=V/U_0
\end{equation} 
where $V$ is the speed of the shape-change \citep{Weymouth2013JFM}. For this maneuver we choose $V=c/\tau$, the average cross-flow velocity of the trailing edge.

\begin{figure}
	\subfloat[Kinematics]{
		\includegraphics[width=0.3\textwidth]{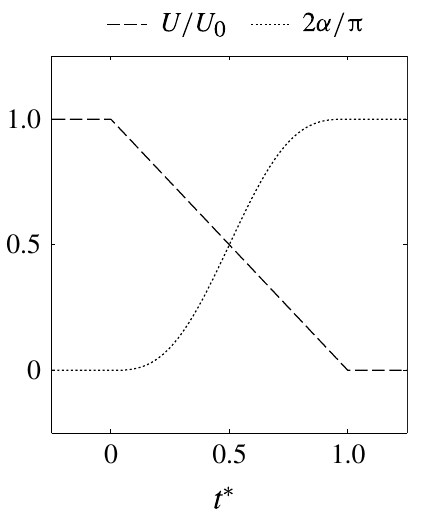}
		}
	\subfloat[Lift coefficient]{
		\includegraphics[width=0.3\textwidth]{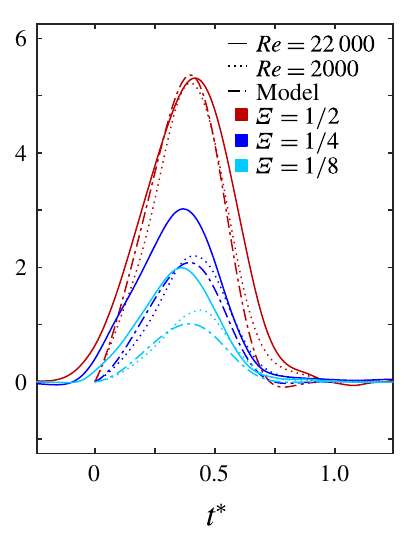}
		}
	\subfloat[Drag coefficient]{
		\includegraphics[width=0.3\textwidth]{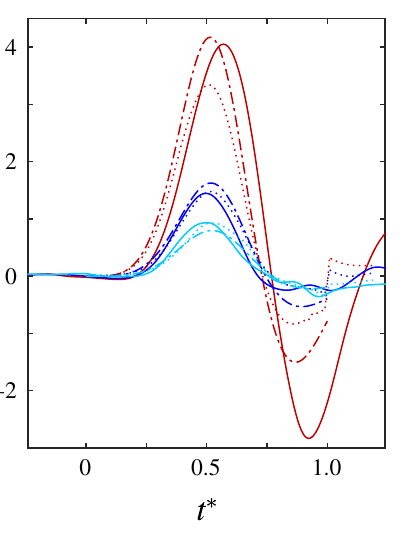}
		}
	\caption{Kinematics and force coefficients (scaled by $U_0$) on a foil with rapidly increasing pitch during deceleration, reproduced from \cite{Polet2015}. The force coefficients from two-dimensional simulations at $Re=2000$, experiments at $Re=22000$, and an inviscid flow model are given over three maneuver speeds.}
	\label{fig:polet}
\end{figure}

Figure~\ref{fig:polet} shows the resulting forces from the simulations, experiments, and an inviscid flow model described in \cite{Polet2015}. Forces increased with increasing shape-change number, and at $\Xi=1/2$ the values are  ten times larger than the lift and drag at the corresponding static pitch angle, which would help birds maintain lift and come to a controlled stop. However, the drag forces are negative at the end of the maneuver which decreases the average stopping force. \cite{Polet2015} postulate that the unwanted thrust generation is due to the prescribed constant rate of deceleration in equation~\ref{eq:eqp}, which does not match the natural fluid-structure interaction in true perching. 

\begin{figure}
	\begin{minipage}{0.47\textwidth}
  		\centering
		\subfloat[$t^*=2.5$]{
			\includegraphics[trim={0 3cm 0 15mm},clip,width=\textwidth]
			{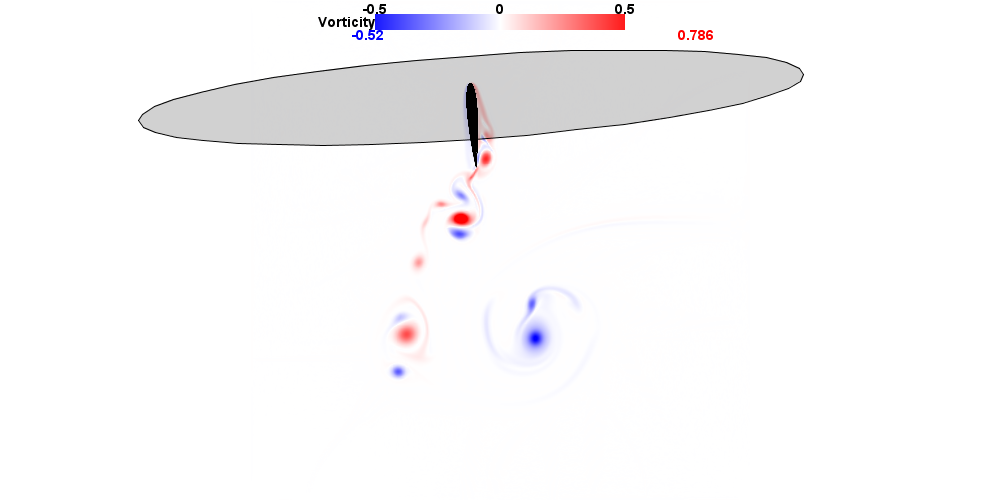}
		} \\
		\subfloat[$t^*=1$]{
			\includegraphics[trim={0 3cm 0 3cm},clip,width=\textwidth]
			{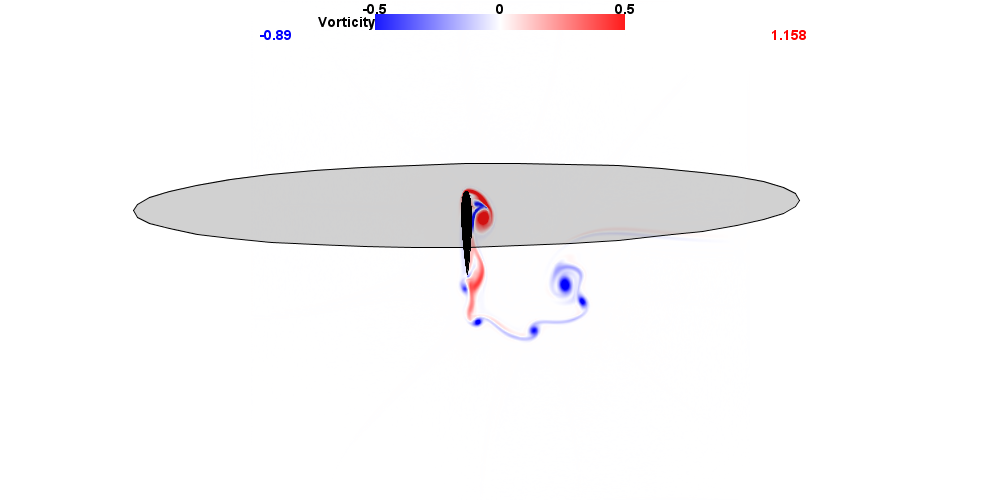}
		} \\
		\subfloat[$t^*=0$]{
			\includegraphics[trim={0 3cm 0 3cm},clip,width=\textwidth]
			{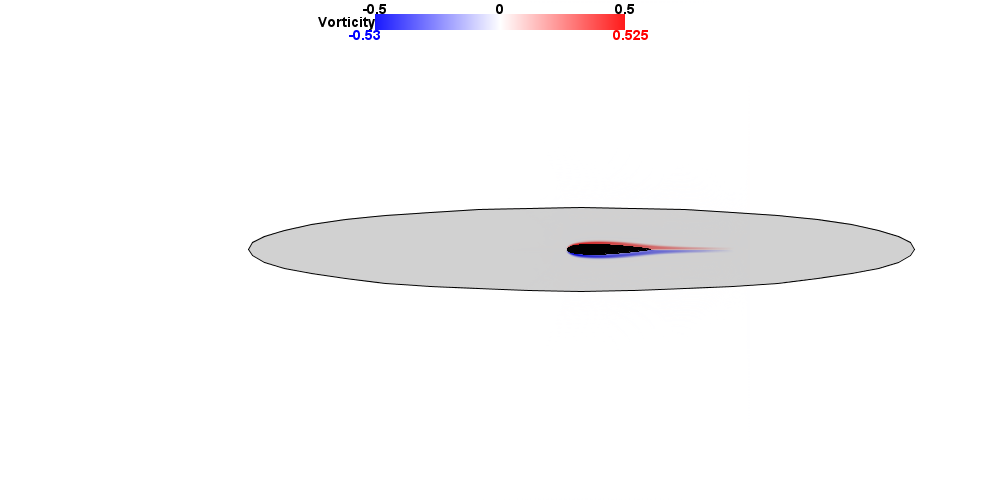}
		}
		\caption{Foil vorticity field for free-running simulations of an ellipsoid undergoing a stopping maneuver by rapidly pitching foils with $\Xi=1/2$, $\theta_{final} = \pi/2$.}
		\label{fig:body_pics}
	\end{minipage}
	\hspace{2mm}
	\begin{minipage}{0.47\textwidth}
  		\centering
		\subfloat[Centroid path]{
			\includegraphics[width=\textwidth]
			{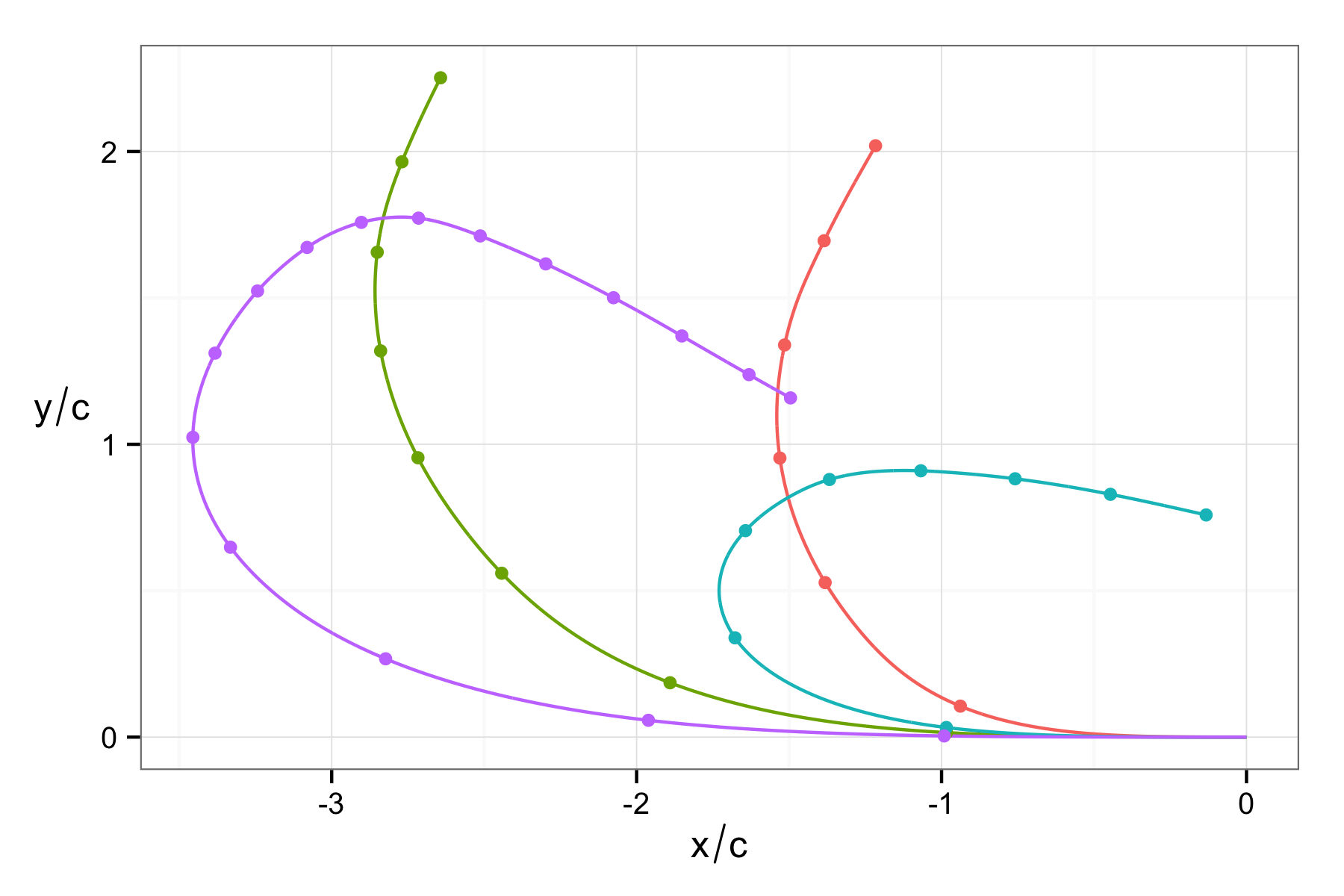}
			} \\
		\subfloat[Drag coefficient]{
			\includegraphics[width=0.5\textwidth]
			{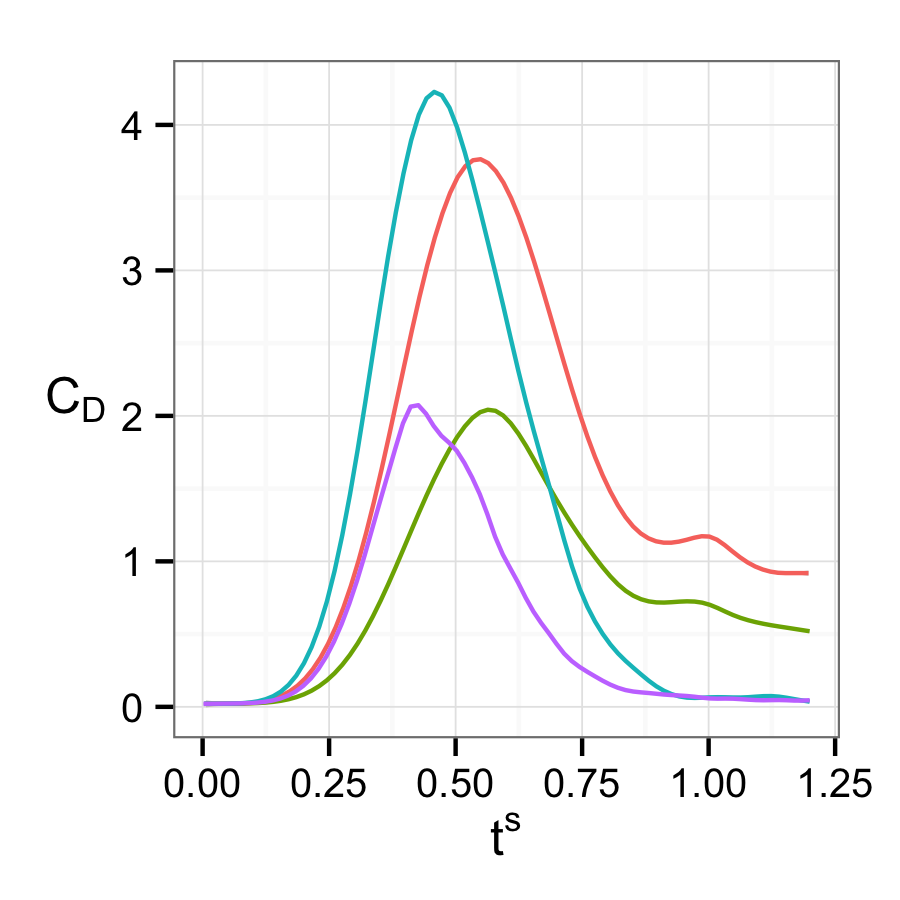}
		}
		\subfloat[Power coefficient]{
			\includegraphics[width=0.5\textwidth]
			{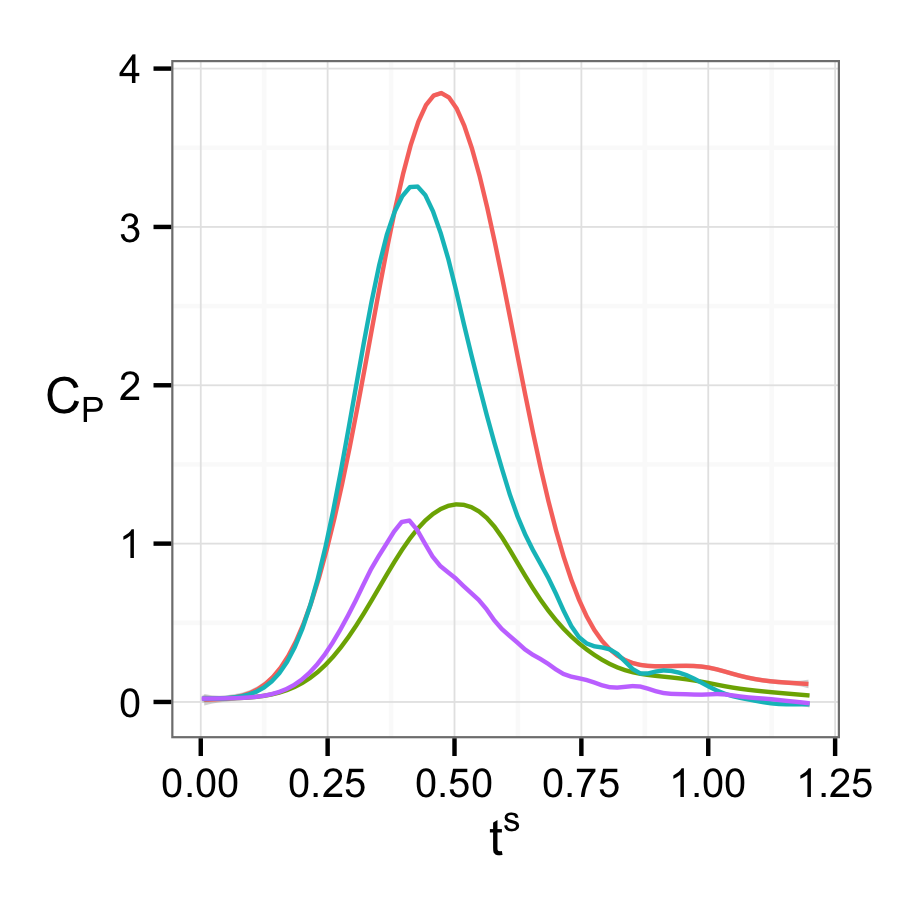}
		}
		\caption{Results for four stopping maneuver cases; 
		$\{\Xi,\theta_{final}\}$ =  
		\textcolor[rgb]{0.5,0,0}{$\{1/2, \pi/2\}$},
		\textcolor[rgb]{0,0.5,0}{$\{1/4, \pi/2\}$},
		\textcolor[rgb]{0,0.5,0.5}{$\{1/2, \pi\}$},
		\textcolor[rgb]{0.5,0,0.5}{$\{1/4, \pi\}$}. 
		Points in (a) show increments of $tU_0/c=1$.}
		\label{fig:body}
	\end{minipage}
\end{figure}

To test this theory and to determine the applicability of pitching foils on maritime vehicles we next carry out free-running simulations of a stopping maneuver. The vehicle is set to be a neutrally buoyant ellipsoid with uniform density, diameter $c$ and length $8c$, Figure~\ref{fig:body_pics}. A NACA0012 foil with span $s$ is mounted on either side of the body center and the pitch relative to the body is given by equation~\ref{eq:turn}. We set $Re=U_0 c/\nu=22000$. The dynamics of the vehicle are modeled as 
\begin{align}
\ddot x = \frac{D-\frac 12 C_x \rho A_x \dot x|\dot x|}{m+m_{xx}},\quad
\ddot y = \frac{L-\frac 12 C_y \rho A_y \dot y|\dot y|}{m+m_{yy}},\quad
\ddot \psi = \frac{M}{I+m_{\psi\psi}}
\end{align}
where $x,y$ are the body centroid location, $\psi$ is the heading, $m$ is the mass, $I$ is the moment of inertia, and $C_a, A_a, m_{aa}$ are the drag coefficient \citep[taken from][]{Hoerner1965}, projected area, and potential flow added-mass in the a-direction. Note that while the fluid forces on the body are modeled analytically, $D,L,M$ are the measured lift drag and moment of the foil in the coupled simulation.

The results of the maneuvering simulations are shown in Figure~ \ref{fig:body}. Increasing the shape-change rate increases the forces, and the peak drag magnitudes are similar to the prescribed deceleration case results in Figure~\ref{fig:polet}. However, the free-running case results in only positive drag force, verifying the Polet et al discussion, and helping the vehicle stop. The resulting trajectories show that the pitch-up maneuver is capable of stopping the body's forward motion in $1.6c$, only 20\% of the body length.

Figure~ \ref{fig:body} also shows two cases where the final pitch has been increased to $\pi$ in equation~\ref{eq:turn}, e.g. the foil keeps pitching until is faces backwards. This motion ensures that it is the dynamic forces responsible for stopping the body - not just bluff-body drag on the sideways foil. The results show the body not only stops, but fully reverses, and does so with relatively little vertical drift.

\section{Size and shape-changing bodies}\label{sec:size}

In contrast to rigid body kinematics, such as flapping, little research has been devoted to explosive size and shape-change despite its prevalence in nature. For example, many animal use ``burst and coast'' gaits when performing maneuvers to reduce the cost of transport by as much as 50\% \citep{Weihs1984,Chung2009}. Extreme shape change is also often used in ``survival'' hydrodynamics, i.e. to help an animal hunt or evade attack where extreme accelerations are required \citep{Triantafyllou2016}. 

In this section we review two series of recent studies on using size-change as a novel form of force generation. Surprisingly, the `ballistic' nature of these novel actuation methods often makes them simpler to implement than the controlled kinematics of the previous section. And advances in soft-robotics are enabling the first tests of these size-and shape changing devices.

\subsection{Span-wise retraction to shed vorticity}

When birds and marine mammals perform ``burst and coast'' maneuvers  they rapidly pull their wings or flippers against their bodies -  causing them to effectively `vanish' from the flow. Classic studies such as \cite{Taylor1953} showed that this sudden disappearance would leave a significant vortex in the fluid, generating large forces. \cite{Wibawa2012} attempted to experimentally and numerically study this vanishing phenomenon by quickly retracting a foil along its span while towing it forward. Retraction is much simpler and less power-consuming than flapping and could be easily used in practical maritime designs. 

The study used a foil with a rectangular planform, square tip, and NACA0012 cross-section. The foil was towed along the tank at $Re=Uc/\nu=14000$ at a $10\deg$ angle of attack and was retracted a distance $1.4c$ with an average speed of $6U$. Experimental results showed that while some circulation was shed, it was less than half of the bound circulation before retracting, and it decayed so quickly that it couldn't be feasibly used to generate maneuvering forces. 

\begin{figure}
	\includegraphics[width=\textwidth]{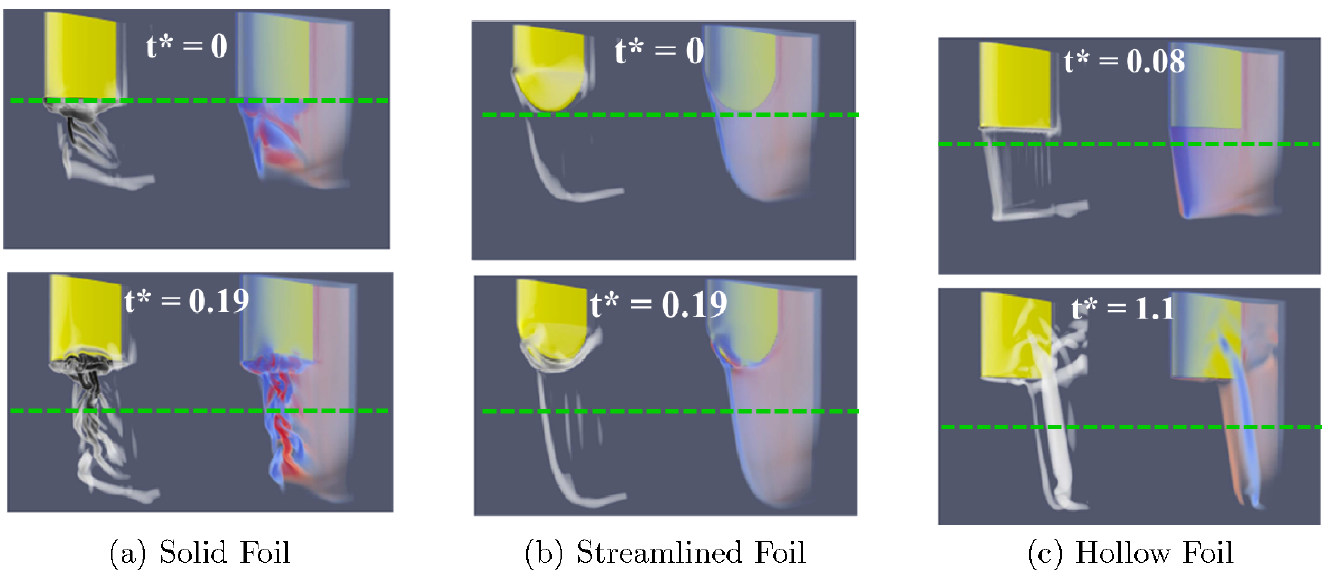}
	\caption{Simulations of the retracting foil at $Re=1000$ for three foil geometries. $t^*=tU/c=0$ is when the foil crosses the PIV plane (green line). The left and right of each panel show $\lambda_2$ and $\omega_z$ iso-surfaces, respectively. Reproduced from \cite{Steele2016b}.}
	\label{fig:steele_sims}
\end{figure}

\begin{figure}
	\centering
	\includegraphics[width=0.8\textwidth]{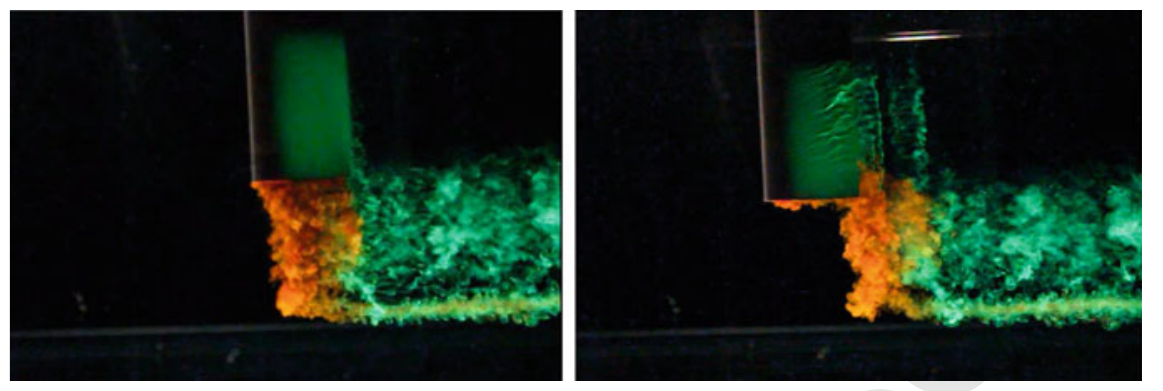}
	\caption{Experiment with dye injection of the retracting open hollow foil at $Re=13700$. The orange dye is from inside the foil, while the green is from the outside. The left image is around $t^*=0.15$, right is around $t^*=1.15$. Reproduced from \cite{Steele2016b}.}
	\label{fig:steele_exp}
\end{figure}

Three-dimensional simulations were performed to visualize the complete flow. Figure~\ref{fig:steele_sims}(a) shows a similar simulation, but run at $Re=1000$ to clarify the vorticity structures. The wake structures were found to be highly complex because the impulsive retraction of the foil generated its own wake, which mixed and disturbed the shedding of the bound vorticity. Again, this limits the amount of useful work that the maneuver can achieve. 

In a follow-up study, \cite{Steele2016b} showed that the shape of the foil geometry can be easily adjusted to achieve different kinds of fluid response. Figure~\ref{fig:steele_sims} shows the result of the same retraction maneuver on two other foil shapes; a foil with a streamlined and rounded wing tip, and a foil which is hollow and open on the wing tip to allow fluid to pass through. Figure~\ref{fig:steele_exp} shows the result of using dye visualization in an experimental test of the retracting hollow foil. The results show that because the hollow foil does need to pull fluid up to fill the wake of its retraction, the vorticity is shed in two large clear vortex structures which could be used to induce dynamic roll moments on trailing control surfaces.

\subsection{Shrinking to recover added-mass energy and cancel drag}

\begin{figure}
	\centering
	\includegraphics[angle=90, width=0.8\textwidth]{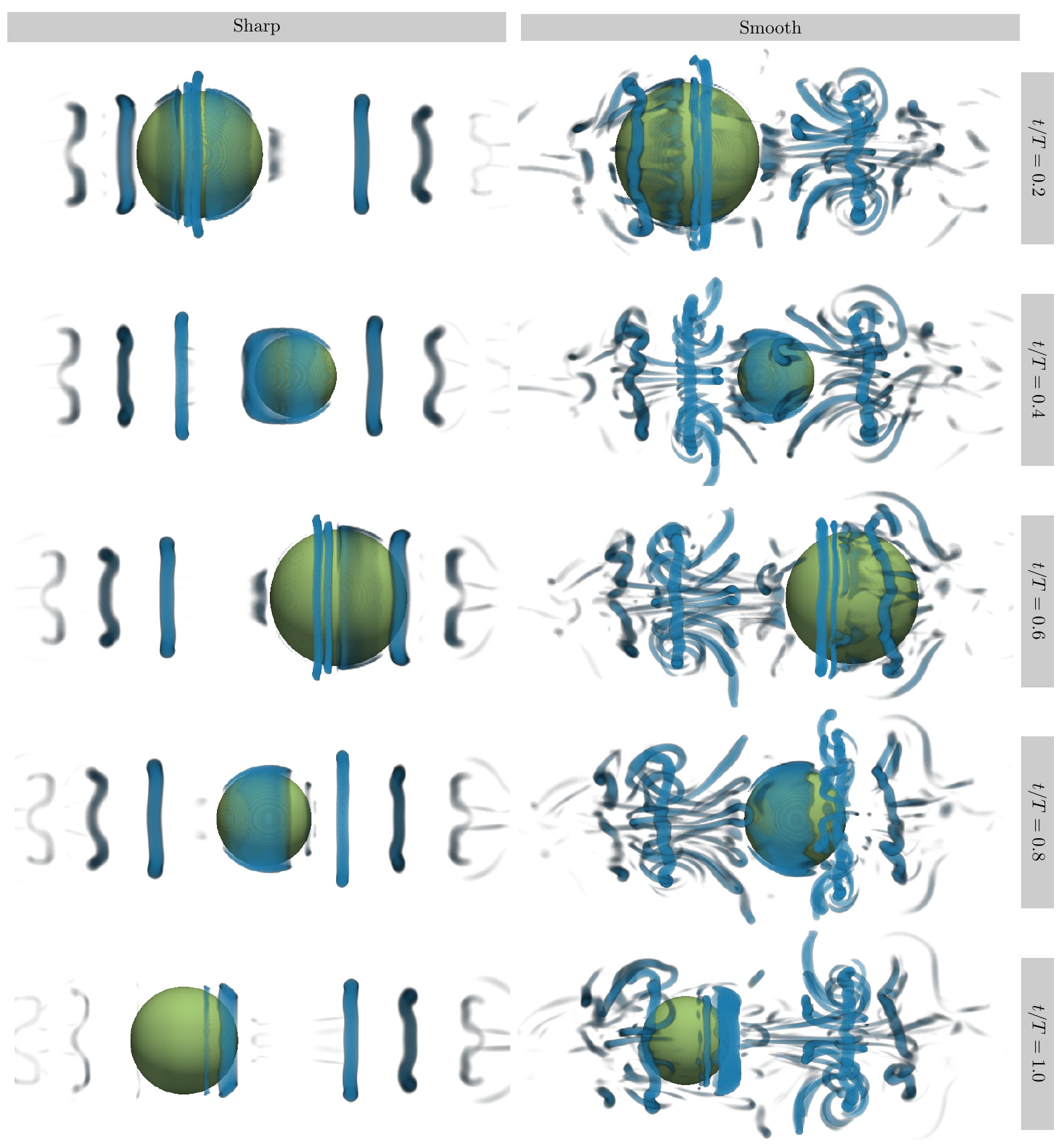}
	\caption{Evolution of the $\lambda_2$ vortex criterion during one oscillation after attainment of zero-damping regime in response to the sharp and smooth radius variations (see Figure~\ref{fig:g-s}). Reproduced from \cite{Giorgio-Serchi2016}.}
\label{fig:lambda2}
\end{figure}

The streamlined foil result in Figure~\ref{fig:steele_sims} is entirely different than that of the hollow foil. Consider the cross-section of the streamlined foil as it retracts through the PIV plane. This is not a `vanishing' body, but a shrinking one. The key difference, as shown in  \cite{Weymouth2012JFM}, is that the shrinking body pulls in fluid to fill the void left by its retraction, while a vanishing hollow body does not. 

In both cases, the reduced size of the body means a corresponding reduction in the fluid added-mass. However, the resulting dynamics of the fluid, and its force on the body could not be more different. For a vanishing body, the surplus fluid kinetic energy goes into the generation of shed vortical structures as shown in Figure~\ref{fig:steele_sims}(c). For a shrinking body, two related effects were found:
\begin{enumerate}
\item The rapid motion of the boundary generates a layer of vorticity which can cancel the boundary layer vorticity for high shape-change numbers. This is demonstrated by the small amount of shed vorticity in Figure~\ref{fig:steele_sims}(b).
\item The cancellation of bound vorticity enables the transfer of the fluid added-mass energy back into the body, resulting in significant instantaneous forces.
\end{enumerate}

In the case of an inviscid fluid, the bound vorticity cancellation is perfect, and the resulting force is simply
\begin{equation}\label{eq:f ma}
F = -\frac{\partial}{\partial t}\left(m_{xx} U \right) = -\dot m_{xx} U - m_{xx} \dot U
\end{equation}
where $\dot m_{xx}$ is the rate of change of the added-mass. The final term is the standard added-mass force due the body acceleration, but the first term is due to the recovery of added-mass energy by the body. For large shape-change numbers $\dot m_{xx} U$ could be sufficient to completely cancel the body drag force.

\begin{figure}[t]
	\centering
	\subfloat[Time history]{
		\includegraphics[width=0.9\textwidth]{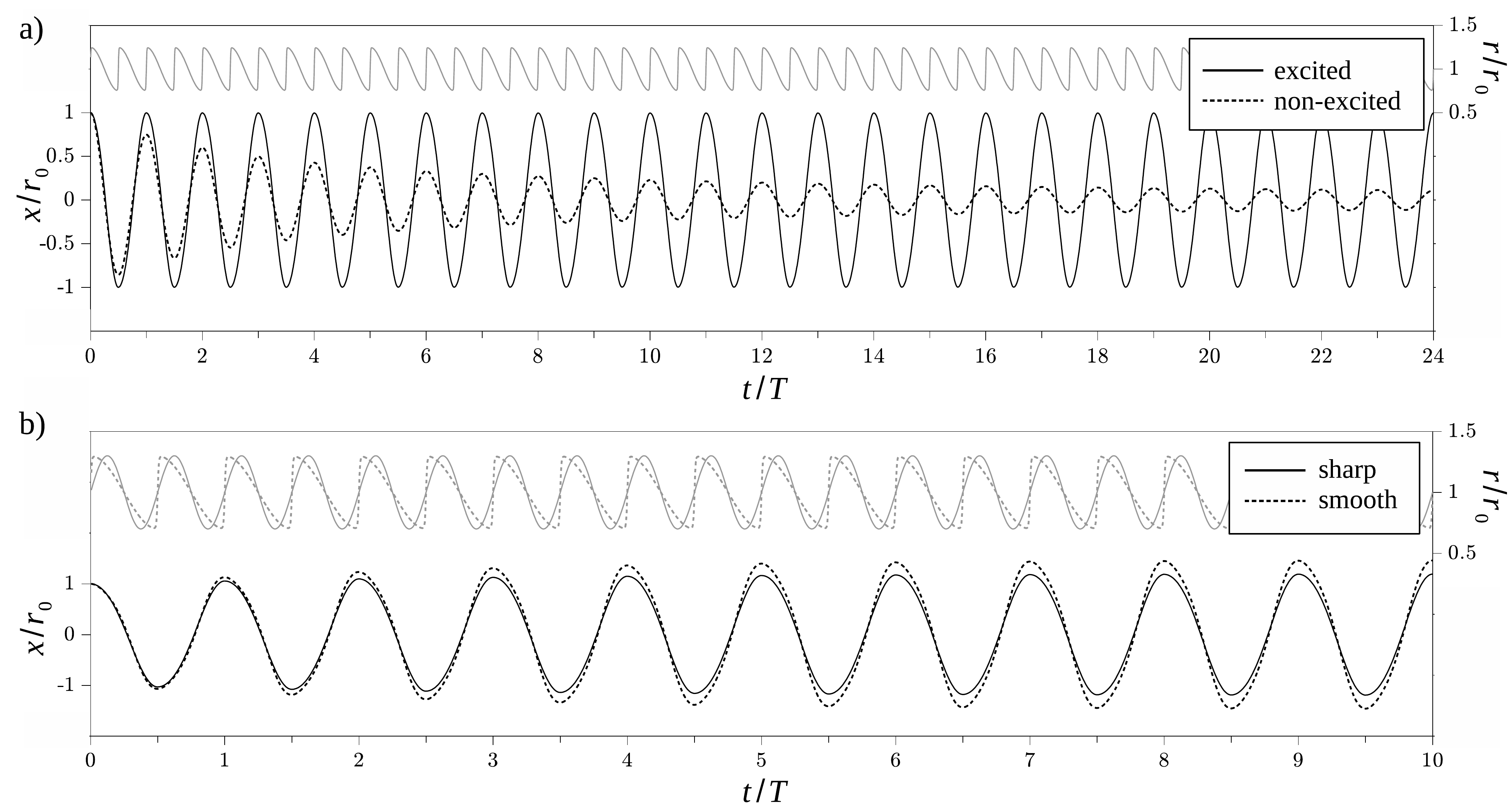}
		} \\
	\subfloat[Frequency dependence]{
		\includegraphics[trim={0 6cm 7cm 6cm}, clip, height=6cm]{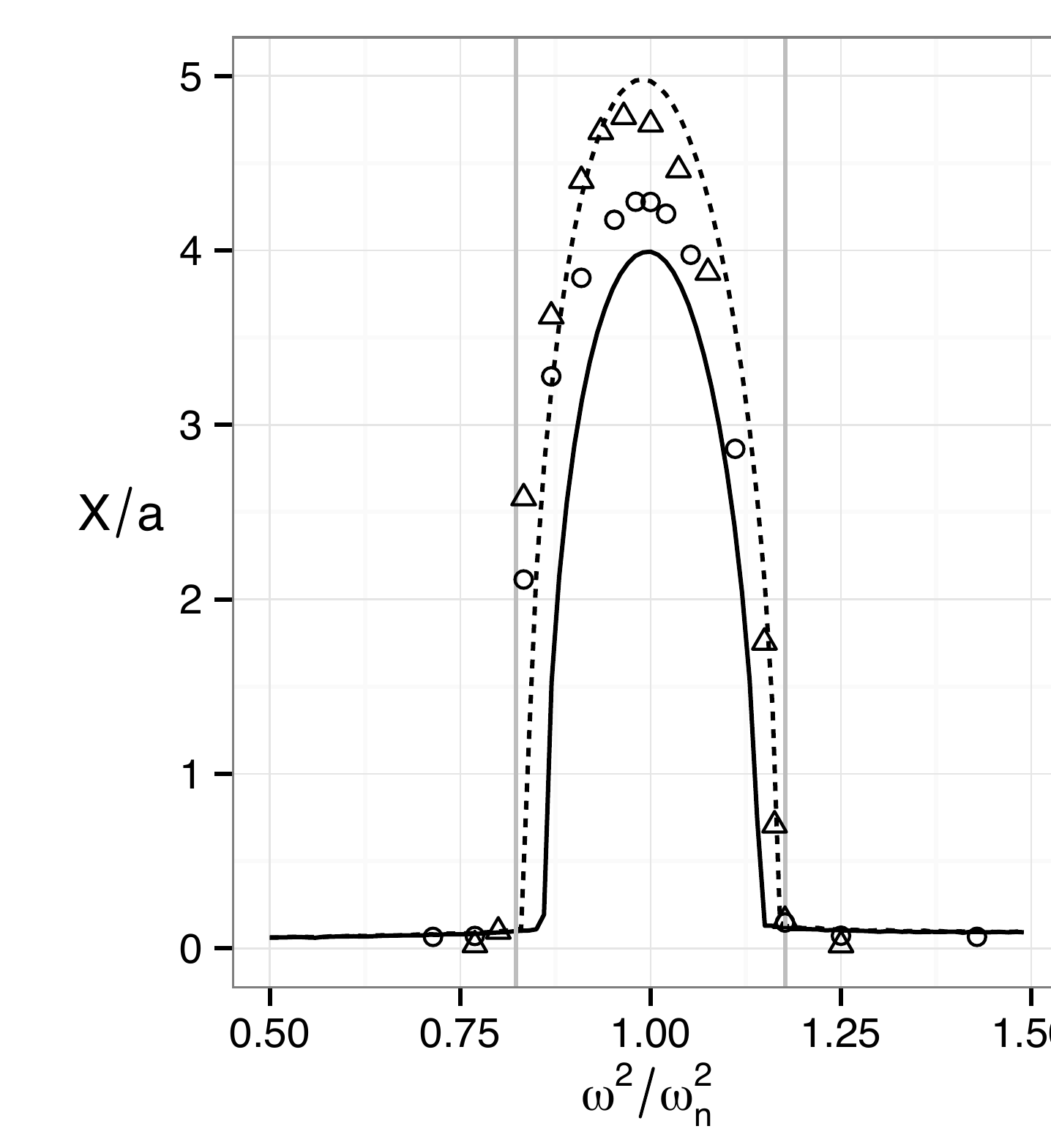}
		}
	\subfloat[Amplitude dependence]{
		\includegraphics[trim={0 6cm 0 6cm}, clip, height=6cm]
		{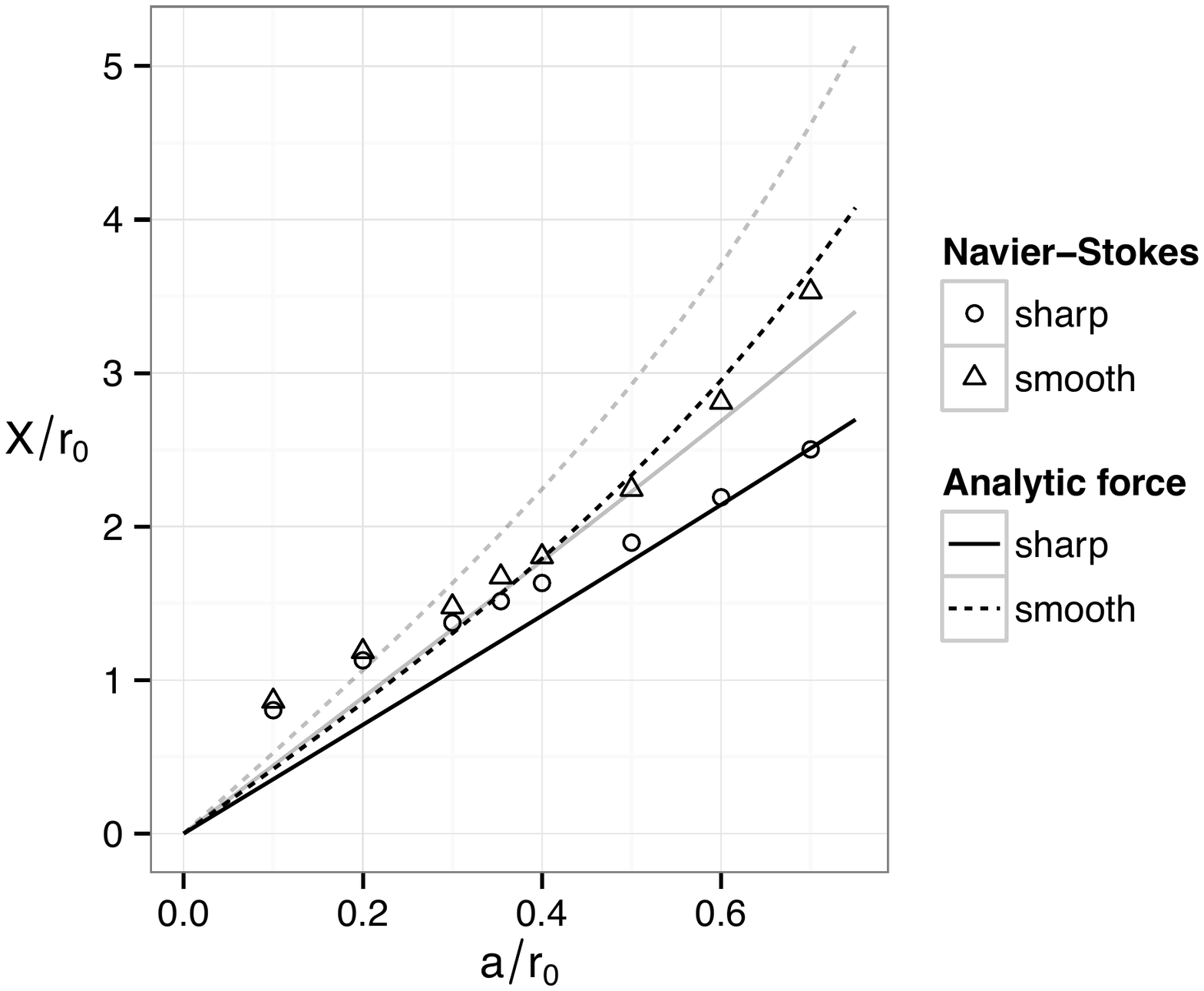}
		}
	\caption{Transverse response $x$ of the oscillating sphere, with and without volume-change excitation. Sharp excitation refers to the `saw tooth` pattern at the top of (a), while smooth excitation is a simple sin wave pattern. The grey lines in (c) assume 100\% efficient added-mass energy recovery, while the dark lines assume $\eta=0.9$. Reproduced from \cite{Giorgio-Serchi2016}.}
	\label{fig:g-s}
\end{figure}

\cite{Giorgio-Serchi2016} used a volume-changing oscillator to test this method of drag cancellation. They simulated the flow on a spherical body with radius $r$ connected to a spring and immersed in water. If this body is released from a large displacement, say $x_0=r$, it will oscillate with a natural frequency $\omega_n$ but the amplitude will quickly decays to nothing due to the drag of the fluid, Figure~\ref{fig:g-s}(a, non-excited). However, if the radius of the sphere changes in time with amplitude $a$, then added-mass energy will transfer back and forth between body and fluid, exciting oscillation, Figure~\ref{fig:lambda2} and \ref{fig:g-s}(a, excited). 

This is called a parametric-oscillator, and just like a child on a swing changing their center of effort, this can lead to sustained large amplitude oscillations if the oscillator is pumped near the natural frequency. But while a swing would work underwater, the amplitude would be tiny due to drag. By shrinking and growing,  the sphere's large bluff body drag force is canceled, enabling oscillation amplitudes up to $X=4.7a$ and $3.5r_0$, Figure~\ref{fig:g-s}. 

Figure~\ref{fig:g-s} also compares the results to an analytic parametric-oscillator model developed in \cite{Giorgio-Serchi2016} using equation \ref{eq:f ma}. While the frequency match is excellent, the model over predicts $X$ for large $a$ because of the imperfect recovery of added-mass energy. Indeed, Figure~\ref{fig:lambda2} shows the simulated flow features large scale vortex shedding - indicating that at least some portion of the energy is spent stirring up the fluid.

To quantify how much energy is wasted, we need to revisit the definition of efficiency. Unlike for an isolated propulsor, the useful work is ill-defined for a self-propelled body. As discussed in \cite{Maertens2015BB} this is because the net force on a steady self-propelled body is zero by definition and the power lost to the environment depends sensitively on the propulsion method. Instead, we must use the quasi-propulsive efficiency
\begin{equation}\label{eq:eqp}
\eta_{QP} = \frac{P_{tow}}{P_{self}}
\end{equation}
where $P_{tow}$ is the power lost to the fluid when \textit{towing the rigid body at its operating condition}, and $P_{self}$ is the power usage measured in the self-propelled test. 

This is, in fact, the standard measure of efficiency used in ship design. In the case of a propeller-driven ship at steady-ahead conditions equation~\ref{eq:eqp} becomes
\begin{equation}
\eta_{QP} = \frac{RU}{Q\omega}
\end{equation}
where the towed resistance $R$ times the speed $U$ is the towed power loss, and the propeller shaft torque $Q$ times the rotation rate $\omega$ is the self-propelled power usage. When using equation~\ref{eq:eqp}, the towed body should be rigid and bare (no propulsor) but otherwise operated at the same conditions as the self-propelled test. 

Applied to the case of the volume-changing and oscillating sphere, we first select a self-propelled case, say $a/r_0=0.35$ and $\omega = \omega_n$ which achieved in $X/a=4.7$ using the smooth profile. We then repeat this case with a rigid sphere towed at the same frequency and amplitude of motion. After using equation~\ref{eq:p} to measure the power used in both cases, the quasi-propulsive efficiency is found to be $\eta_{QP}=0.91$. This was found to be a representative value for the resonant smooth profile cases.\footnote{ Note that the power transfer during sharp inflation is infinite, making this a rather poor choice for energy efficiency. Even use a slightly smoothed profile, the extreme magnitude of the power peaks made computing a meaningful average impossible.} Using this value, the analytic prediction can be corrected and agrees well with experiments, Figure~\ref{fig:g-s}.

If drag cancellation with 90\% efficiency seems too good to be true, it may be explained (or perhaps rationalized) by considering that the growing and shrinking of a shape in water induces a completely irrotational fluid motion. Unlike the rotation of a propeller or flapping of a foil then, an inflate-deflate cycle is perfectly reversible, resulting in a zero net transfer of energy to the fluid over the cycle. As the maturing field of soft robotics enables designs with highly deformable parts \citep{Giorgio-Serchi2013}, such efficiencies may be soon be realized experimentally. 

\subsection{Deflating to power an ultra-fast start}

Cephalopods, such as the squid and octopus, greatly increase their size by filling with water, before ejecting the water in a propulsive jet, reducing their size and helping them make a quick escape \citep{Huffard2006}. As a final example of biologically inspired-force production, we review a series of studies that investigated a jet-propelled shrinking vehicle as a model of this system both analytically and experimentally.

\cite{Weymouth2013JFM} consider three types of jet-propelled bodies; a rocket in the vacuum of space, a rigid 5:1 ellipsoidal torpedo in water, and an octopus-like vehicle which shrinks from a sphere to a 5:1 ellipsoid as it jets. The acceleration of all three is governed by the simple equation
\begin{equation}
\ddot x = \frac{F-\dot m U_J}{m}=\frac{\sum F}{m}
\end{equation}
where $\sum F$ is the total force, which is the fluid force $F$ plus the jet thrust $T_J=-\dot m U_J$, $-\dot m$ is the rate of mass loss and $U_J$ is the jet exit velocity.

\begin{figure}
	\includegraphics[width=\textwidth]{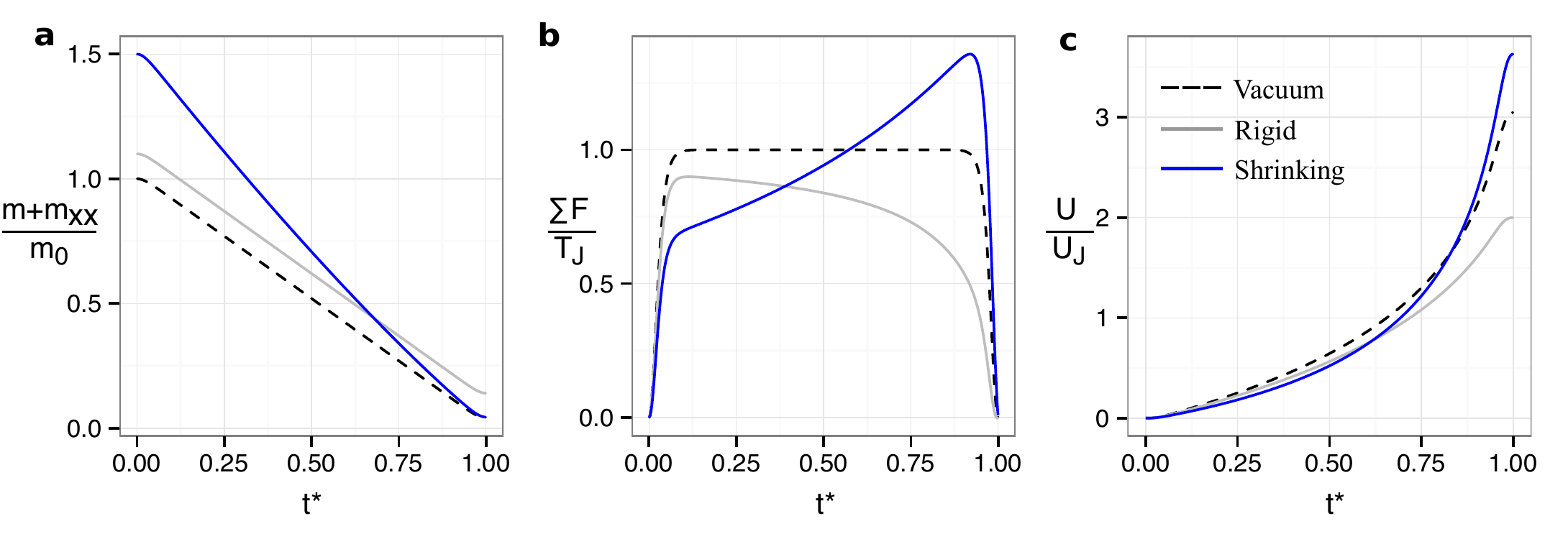}
	\caption{Comparison of three jet-propelled rocket fast-start maneuvers using equation~\ref{eq:f ma} to model the fluid reaction force.}
	\label{fig:rocket_theory}
\end{figure}

Figure~\ref{fig:rocket_theory} shows the results for all the three cases when jetting from rest until 96\% of the initial mass $m_0$ has been expelled, keeping $U_J$ constant for the majority of the maneuver. 
\begin{itemize}
\item In a vacuum, $F=0$ and the net force $\sum F$ equals $T_J$. The rocket accelerates at an increasing rate due to decreased inertia, accelerating far beyond the jet velocity.
\item If we model the fluid reaction force on the rigid torpedo with equation~\ref{eq:f ma}, then the body experiences no drag, but will have an ever increasing added-mass force such that $\sum F<<T_J$. In essence, the torpedo's added mass is an additional payload which it never sheds, limiting its acceleration. 
\item The octopus-like vehicle starts as a sphere, meaning its inertia is initially 50\% greater than the rocket in space. However, unlike the rigid torpedo, this is not payload - it is additional propellant! As the body shrinks, the added-mass energy is recovered in the form of thrust by equation~\ref{eq:f ma}. In the second half of the maneuver, when the inertia is reduced, this results in $\sum F>> T_J$. 
\end{itemize}
The final result being that the octopus-like body accelerates to speeds above $3.5U_J$, much faster than the rigid torpedo, and even faster than a rocket in the vacuum of space.

As discussed above, the successful recovery of added-mass energy requires that the energy is not lost to shed vorticity. \cite{Weymouth2015} studied this process and developed an analytic parameter to predict the recovery efficiency. As the octopus-like vehicle shrinks, it induces a normal velocity which draws in the boundary layer fluid. If this inward velocity is strong enough to overcome the diffusion of the boundary layer, then the vorticity can be annihilated and the flow energy recovered. \cite{Weymouth2015} liken this to the application of suction on a rigid boundary layer. In analogy to a suction parameter, they define a shrinking parameter
\begin{equation}
\sigma^* = V\sqrt{\frac L{U\nu}} = \Xi \sqrt{Re}
\end{equation}
where $V$ is the cross-flow velocity of the deforming body, in this case the rate of change of the minor-axis radius. This modification of the shape-change number includes the rate of boundary layer diffusion, and axis-symmetric boundary layer theory suggests that $\sigma^* > 9 $ should be a thresh-hold value for delayed separation and energy recovery. Note that this threshold is easier to achieve at larger Reynolds numbers, and therefore large body-sizes.

\begin{figure}
	\includegraphics[width=\textwidth]{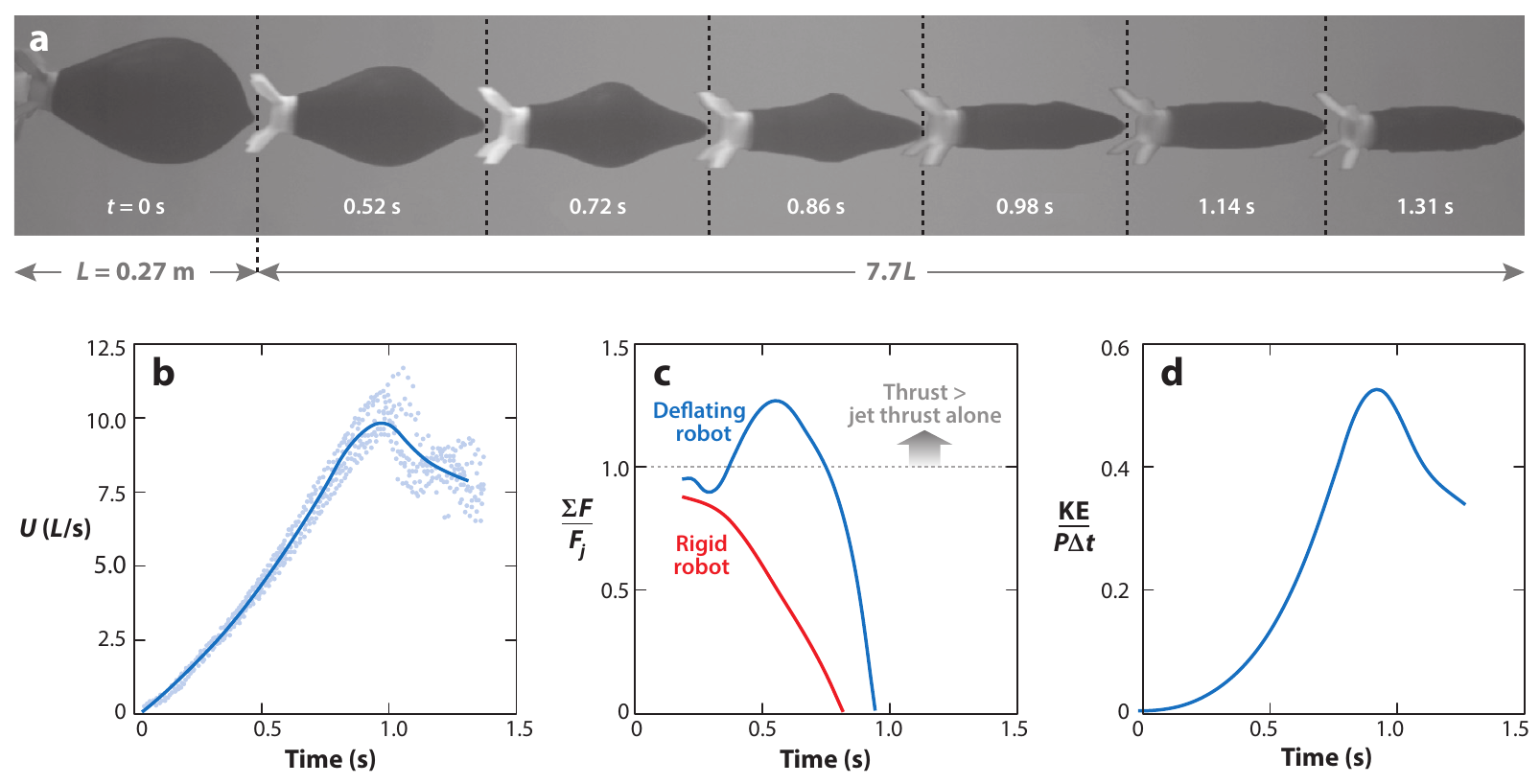}
	\caption{Results of the self-propelled test of the octopus-inspired vehicle. Reproduced from \cite{Triantafyllou2016}, where $F_j=T_J$ is the jet thrust.}
	\label{fig:arfm}
\end{figure}

Based on this, \cite{Weymouth2015} designed a prototype soft robotics vehicle to maximize $\sigma^*$ during a jet-propelled fast start maneuver. The octopus-inspired vehicle consists of a rigid neutrally buoyant skeleton with an elastic membrane stretched around it to form the outer hull, Figure~\ref{fig:arfm}(a). As with the mantle of the octopus, this membrane can be inflated, giving it an initially bluff shape and storing sufficient energy to power its escape. The fully deflated hull shape is approximately a 5:1 ellipsoid, and is sufficiently streamlined to allow the body to coast dozens of body lengths. The body length is $L=26cm$ and the volume when fully deflated is $1030 cm^3$, so the `payload' mass accelerated by the maneuver is $m_f = 1.03~kg$.

Once inflated, the robot is released from a mount allowing it to accelerate forward in open water under its own power. The resulting fast-start maneuver performance is measured using high-speed cameras at 150 frames/second. Figure~\ref{fig:arfm} shows the rapid acceleration and deflation of the shrinking robot from a self-propelled run. The velocity peaks above $10L/s$ or $2.7m/s$ around $t=0.95s$ after release. Based on this and scale of deformation of the body we have $\Xi_{ave}=1/24$, $Re_{ave}=350000$ and the shrinking parameter is $\sigma^*>77$ throughout the maneuver, well above the threshold. 

This high shrinking parameter indicates we should have efficient energy recovery. To quantify this the outline of the body is measured from the images to determine the mass, mass flux, net force ($m\ddot x$), and jet thrust ($-\dot m U_J$) during the maneuver. Figure~\ref{fig:arfm} shows that the peak net force is 30\% greater than $T_J$, similar to the analytic predictions. 

We can also measure the payload kinetic energy and the integrated power delivered by the jet:
\begin{equation}
KE = \frac 12 m_f U^2, \quad
P\Delta t = \frac 12 \int_0^\tau \dot m U_J^2 dt
\end{equation}
Figure~\ref{fig:arfm} shows the ratio of these values, which peaks at $53\%$. This is on par with the theoretical propulsive efficiency of rocket accelerating from rest in a vacuum, which peaks at $65\%$ \citep{Ivey1947}. 

However, this is \textit{not} the quasi-propulsive efficiency of the prototype. The integrated $P_{tow}$ is the change in kinetic energy \textbf{plus} the integral of $RU$ when towing the deflated body through the same maneuver. Using the conservative values $C_D=0.05$ and $m_{xx}=m_f/10$ for the deflated shape gives a quasi-propulsive efficiency of $\eta_{QP}=68\%$, better than a rocket in space.

\section{Discussion and Conclusions}

Vorticity generation is the key to all fluid force generation. It is text-book knowledge that increasing the speed of a body will generally generate more vorticity and increase the force. Slightly less well known is that added-mass in a viscous fluid is also based on vorticity generation on the body surface, making this a uniting theme in fluid dynamics \citep{Wu2007book}. 

In this context, one characteristic stands out in the biological-inspired studies above: 
\begin{quote}
	\textit{Unsteady biologically-based propulsors optimize the generation of vorticity by coordinating their kinematics and shape-change with the state of the flow.}
\end{quote}

The additional degrees of freedom in  biologically-based systems gives them the potential to generate vorticity when and where it will be most useful, and this can be utilized to efficiently produce large forces for maritime applications.

\begin{itemize}
\item In the case of tandem flapping foils, proper phase and distance gaps between the foils enable positive interference to double the thrust on the back fol, or to reduce the variation in lift and thrust. 
\item A foil pitched-up rapidly is capable of generating large vorticity if the shape-change number $\Xi$ is increased, and can bring a streamlined body to a complete stop in 20\% of its length. 
\item Spanwise-retraction of a hollow foil minimizes the generation of new vorticity, freeing the bound vorticity to do other useful work.
\item On the other hand, retracting a foil with a streamlined planform generates opposite-sign vorticity on the boundary, annihilating the bound vorticity.
\item This annihilation enables a body to recover the fluid's added-mass kinetic energy in the form of a large unsteady force. If timed with the natural frequency, this can be used to cancel drag on a size-changing sphere with 91\% efficiency.
\item Finally, by treating the added-mass as additional propellant, stored up initially and released throughout, a shrinking underwater vehicle can achieve an ultra-fast start.
\end{itemize} 

This recovery of fluid energy in the form of thrust is especially interesting, and occurs readily as long as its shrinking rate $\sigma^*$ overcomes viscous diffusion. As this number increases with Reynolds number, even greater quasi-propulsive efficiency may soon be realized experimentally. 

\section*{Acknowledgements}

This work was performed in collaboration with excellent research groups world-wide; including Michael Triantafyllou's group at MIT, David Rival's group at Queens University, Brenden Epp's group at Dartmouth University, and Bharathram Ganapathisubramani's group at University of Southampton. 

\bibliographystyle{plainnat}
\bibliography{../../complete.bib}

\begin{thebibliography}{28}
\providecommand{\natexlab}[1]{#1}
\providecommand{\url}[1]{\texttt{#1}}
\expandafter\ifx\csname urlstyle\endcsname\relax
  \providecommand{\doi}[1]{doi: #1}\else
  \providecommand{\doi}{doi: \begingroup \urlstyle{rm}\Url}\fi

\bibitem[Carruthers et~al.(2007)Carruthers, Thomas, and Taylor]{Carruthers2007}
Anna~C Carruthers, Adrian~LR Thomas, and Graham~K Taylor.
\newblock Automatic aeroelastic devices in the wings of a steppe eagle aquila
  nipalensis.
\newblock \emph{Journal of Experimental Biology}, 210\penalty0 (23):\penalty0
  4136--4149, 2007.

\bibitem[Chung(2009)]{Chung2009}
MH~Chung.
\newblock On burst-and-coast swimming performance in fish-like locomotion.
\newblock \emph{Bioinspiration \& biomimetics}, 4\penalty0 (3):\penalty0
  036001, 2009.

\bibitem[Epps et~al.(2016)Epps, Muscutt, Roesler, Weymouth, and
  Ganapathisubramani]{Epps2016}
Brenden~P Epps, Luke~E Muscutt, Bernard~T Roesler, Gabriel~D Weymouth, and
  Bharathram Ganapathisubramani.
\newblock On the interfoil spacing and phase lag of tandem flapping foil
  propulsors.
\newblock \emph{Journal of Ship Production and Design}, 2016.

\bibitem[Giorgio-Serchi and Weymouth(2016)]{Giorgio-Serchi2016}
F.~Giorgio-Serchi and G.~D. Weymouth.
\newblock Drag cancellation by added-mass pumping.
\newblock \emph{Journal of Fluid Mechanics}, 798, Jun 2016.
\newblock \doi{10.1017/jfm.2016.353}.
\newblock URL \url{http://dx.doi.org/10.1017/jfm.2016.353}.

\bibitem[Giorgio-Serchi et~al.(2013)Giorgio-Serchi, Arienti, and
  Laschi]{Giorgio-Serchi2013}
Francesco Giorgio-Serchi, Andrea Arienti, and Cecilia Laschi.
\newblock Biomimetic vortex propulsion: toward the new paradigm of soft
  unmanned underwater vehicles.
\newblock \emph{IEEE/ASME Transactions On Mechatronics}, 18\penalty0
  (2):\penalty0 484--493, 2013.

\bibitem[Hoerner(1965)]{Hoerner1965}
Sighard~F Hoerner.
\newblock \emph{Fluid-dynamic drag: practical information on aerodynamic drag
  and hydrodynamic resistance}.
\newblock Hoerner Fluid Dynamics Midland Park, NJ, USA, 1965.

\bibitem[Huffard(2006)]{Huffard2006}
C.~L. Huffard.
\newblock Locomotion by abdopus aculeatus (cephalopoda: Octopodidae): walking
  the line between primary and secondary defenses.
\newblock \emph{J. Expl. Biol.}, 209:\penalty0 3697--3707, 2006.

\bibitem[Ivey(1947)]{Ivey1947}
H~Reese Ivey.
\newblock Letter to editor.
\newblock \emph{Journal of the Aeronautical Sciences}, 14\penalty0
  (8):\penalty0 450--450, 1947.
\newblock \doi{10.2514/8.1409}.
\newblock URL \url{http://dx.doi.org/10.2514/8.1409}.

\bibitem[Lighthill(1960)]{Lighthill1960}
MJ~Lighthill.
\newblock Note on the swimming of slender fish.
\newblock \emph{Journal of fluid Mechanics}, 9\penalty0 (02):\penalty0
  305--317, 1960.

\bibitem[Maertens et~al.(2015)Maertens, Triantafyllou, and Yue]{Maertens2015BB}
AP~Maertens, MS~Triantafyllou, and DKP Yue.
\newblock Efficiency of fish propulsion.
\newblock \emph{Bioinspiration \& Biomimetics}, 10\penalty0 (4), 2015.

\bibitem[Maertens and Weymouth(2015)]{Maertens2015}
Audrey~P Maertens and Gabriel~D Weymouth.
\newblock Accurate cartesian-grid simulations of near-body flows at
  intermediate reynolds numbers.
\newblock \emph{Computer Methods in Applied Mechanics and Engineering},
  283:\penalty0 106--129, 2015.

\bibitem[McCurdy et~al.(1941)]{Mccurdy1941}
Edward McCurdy et~al.
\newblock \emph{The notebooks of Leonardo da Vinci}.
\newblock Garden City publishing co., inc., 1941.

\bibitem[Polet et~al.(2015)Polet, Rival, and Weymouth]{Polet2015}
Delyle~T. Polet, David~E. Rival, and Gabriel~D. Weymouth.
\newblock Unteady dynamics of rapid perching manoeuvres.
\newblock \emph{Journal of Fluid Mechanics}, 767:\penalty0 323--341, 2015.
\newblock \doi{10.1017/jfm.2015.61}.

\bibitem[Provini et~al.(2014)Provini, Tobalske, Crandell, and
  Abourachid]{Provini2014}
P.~Provini, B.~W. Tobalske, K.~E. Crandell, and A.~Abourachid.
\newblock Transition from wing to leg forces during landing in birds.
\newblock \emph{J. Exp. Biol.}, 2014.
\newblock \doi{10.1242/jeb.104588}.
\newblock URL
  \url{http://jeb.biologists.org/content/early/2014/05/08/jeb.104588.abstract}.

\bibitem[Read et~al.(2003)Read, Hover, and Triantafyllou]{Read2003}
Douglas~A Read, FS~Hover, and MS~Triantafyllou.
\newblock Forces on oscillating foils for propulsion and maneuvering.
\newblock \emph{Journal of Fluids and Structures}, 17\penalty0 (1):\penalty0
  163--183, 2003.

\bibitem[Simpson et~al.(2008)Simpson, Licht, Hover, and
  Triantafyllou]{Simpson2008}
Bradley~J Simpson, Stephen Licht, Franz~S Hover, and Michael~S Triantafyllou.
\newblock Energy extraction through flapping foils.
\newblock In \emph{ASME 2008 27th International Conference on Offshore
  Mechanics and Arctic Engineering}, pages 389--395. American Society of
  Mechanical Engineers, 2008.

\bibitem[Steele et~al.(2016)Steele, Dahl, Weymouth, and Triantafy]{Steele2016b}
S.C. Steele, J.M. Dahl, G.D. Weymouth, and M.S. Triantafy.
\newblock Shape of retracting foils that model morphing bodies controls shed
  energy and wake structure.
\newblock \emph{Journal of Fluid Mechanics}, to appear, 2016.

\bibitem[Taylor(1953)]{Taylor1953}
GI~Taylor.
\newblock Formation of a vortex ring by giving an impulse to a circular disk
  and then dissolving it away.
\newblock \emph{Journal of Applied Physics}, 24\penalty0 (1):\penalty0
  104--104, 1953.

\bibitem[Triantafyllou et~al.(2000)Triantafyllou, Triantafyllou, and
  Yue]{Triantafyllou2000}
M.~S. Triantafyllou, G.~S. Triantafyllou, and D.~K.~P. Yue.
\newblock Hydrodynamics of fishlike swimming.
\newblock \emph{Annual Review of Fluid Mechanics}, 32\penalty0 (1):\penalty0
  33--53, 2000.
\newblock \doi{10.1146/annurev.fluid.32.1.33}.
\newblock URL
  \url{http://www.annualreviews.org/doi/abs/10.1146/annurev.fluid.32.1.33}.

\bibitem[Triantafyllou et~al.(2016)Triantafyllou, Weymouth, and
  Miao]{Triantafyllou2016}
Michael~S. Triantafyllou, Gabriel~D. Weymouth, and Jianmin Miao.
\newblock Biomimetic survival hydrodynamics and flow sensing.
\newblock \emph{Annual Review of Fluid Mechanics}, 48\penalty0 (1):\penalty0
  null, 2016.
\newblock \doi{10.1146/annurev-fluid-122414-034329}.
\newblock URL \url{http://dx.doi.org/10.1146/annurev-fluid-122414-034329}.

\bibitem[Triantafyllou et~al.(1994)Triantafyllou, Grosenbaugh, and
  Gopalkrishnan]{Triantafyllou1994}
MS~Triantafyllou, MA~Grosenbaugh, and R~Gopalkrishnan.
\newblock Vortex-induced vibrations in a sheared flow: a new predictive method.
\newblock Technical report, DTIC Document, 1994.

\bibitem[Weihs and Webb(1984)]{Weihs1984}
D~Weihs and Paul~W Webb.
\newblock Optimal avoidance and evasion tactics in predator-prey interactions.
\newblock \emph{Journal of Theoretical Biology}, 106\penalty0 (2):\penalty0
  189--206, 1984.

\bibitem[Weymouth(2015)]{Weymouth2015b}
G~D Weymouth.
\newblock Lily pad: Towards real-time interactive computational fluid dynamics.
\newblock In Volker Bertram \& Emilio~F. Campana, editor, \emph{18th Numerical
  Towing Tank Symposium}, Cortona Italy, 28-30 September 2015.

\bibitem[Weymouth and Triantafyllou(2012)]{Weymouth2012JFM}
G~D Weymouth and M~S Triantafyllou.
\newblock Global vorticity shedding for a shrinking cylinder.
\newblock \emph{Journal of Fluid Mechanics}, 702:\penalty0 470--487, 2012.

\bibitem[Weymouth et~al.(2015)Weymouth, Subramaniam, and
  Triantafyllou]{Weymouth2015}
G~D Weymouth, V~Subramaniam, and M~S Triantafyllou.
\newblock Ultra-fast escape maneuver of an octopus-inspired robot.
\newblock \emph{Bioinspir. Biomim.}, 10\penalty0 (1):\penalty0 016016, feb
  2015.
\newblock \doi{10.1088/1748-3190/10/1/016016}.
\newblock URL \url{http://dx.doi.org/10.1088/1748-3190/10/1/016016}.

\bibitem[Weymouth and Triantafyllou(2013)]{Weymouth2013JFM}
Gabriel~D. Weymouth and M.~S. Triantafyllou.
\newblock Ultra-fast escape of a deformable jet-propelled body.
\newblock \emph{Journal of Fluid Mechanics}, 721:\penalty0 367--385, 2013.

\bibitem[Wibawa et~al.(2012)Wibawa, Steele, Dahl, Rival, Weymouth, and
  Triantafyllou]{Wibawa2012}
M.~S. Wibawa, S.~C. Steele, J.~M. Dahl, D.~E. Rival, G.~D. Weymouth, and M.~S.
  Triantafyllou.
\newblock Global vorticity shedding for a vanishing wing.
\newblock \emph{Journal of Fluid Mechanics}, 695:\penalty0 112--134, 2012.

\bibitem[Wu et~al.(2007)Wu, Ma, and Zhou]{Wu2007book}
Jie-Zhi Wu, Hui-Yang Ma, and Ming-De Zhou.
\newblock \emph{Vorticity and vortex dynamics}.
\newblock Springer Science \& Business Media, 2007.

\end{thebibliography}
\end{document}